\documentclass[aps,pra,graphicx,9pt,twocolumn,superscriptaddress,showkeys,showpacs]{revtex4-2}
\usepackage[colorlinks,linkcolor=blue,urlcolor=blue,anchorcolor=blue,citecolor=blue]{hyperref}
\usepackage{amsmath}
\UseRawInputEncoding
\usepackage{graphicx}
\usepackage{dcolumn}
\usepackage{mathrsfs}
\usepackage{amssymb}
\usepackage{amsfonts,multirow}
\usepackage{bm,physics}
\usepackage[defaultcolor=red,final]{changes}

\begin{document}

\title{Non-adiabatic holonomic quantum operations in continuous variable systems}

\author{Hao-Long Zhang}
\author{Yi-Hao Kang}
\author{Fan Wu}

\author{Zhen-Biao Yang}
\email{zbyang@fzu.edu.cn}

\author{Shi-Biao Zheng}
\email{t96034@fzu.edu.cn}

\affiliation{%
 Fujian Key Laboratory of Quantum Information and Quantum Optics, College of Physics and Information Engineering, Fuzhou University, Fuzhou,Fujian 350116, China 
}%

\date{\today}
\begin{abstract}

    Quantum operations by utilizing the underlying geometric phases produced in physical systems are favoured due to its potential robustness. When a system in a non-degenerate eigenstate undergoes an adiabatically cyclic evolution dominated by its Hamiltonian, it will get a geometric phase, referred to as the Berry Phase. While a non-adiabatically cyclic evolution produces an Aharonov-Anandan geometric phase. The two types of Abelian geometric phases are extended to the non-Abelian cases, where the phase factors become matrix-valued and the transformations associated with different loops are non-commutable. Abelian and non-Abelian (holonomic) operations are prevalent in discrete variable systems, whose limited (say, two) energy levels, form the qubit. While their developments in continuous systems have also been investigated, mainly due to that, bosonic modes (in, such as, cat states) with large Hilbert spaces, provide potential advantages in fault-tolerant quantum computation. Here we propose a feasible scheme to realize non-adiabatic holonomic quantum logic operations in continuous variable systems with cat codes. We construct arbitrary single-qubit (two-qubit) gates with the combination of single- and two-photon drivings applied to a Kerr Parametric Oscillator (KPO) (the coupled KPOs). Our scheme relaxes the requirements of the previously proposed adiabatic holonomic protocol dependent on long operation time, and the non-adiabatic Abelian ones relying on a slight cat size or an ancilla qutrit.
\end{abstract}


\maketitle


\section{Introduction}
    The concept of geometric phases was first proposed by Berry \cite{RN17}, who discovered that a quantum system in a non-degenerate eigenstate of an adiabatically and cyclically changed Hamiltonian will pick up a path-dependent phase, in addition to the dynamical one. This phase only depends on the global geometric feature of the loop traversed in parameter space, and hence is insensitive to fluctuations of the control parameters \cite{RN12}. This feature is favorable for implementation of quantum logic gates theoretically \cite{Nat4076802,RN6} and experimentally \cite{RN33, RN36}. However, the long operation time imposed by the adiabatic condition represents a disadvantage in view of decoherence \cite{PhysRevLett.104.120401, RN27}. To overcome this problem, one treatment is to design driving algorithms or optimized techniques to speed up this process \cite{RN40,RN43,RN45,RN46,RN296}. The other is non-adiabatic geometric quantum computation protocols, which are based on the non-adiabatic geometric phase proposed by Aharonov and Anandan \cite{RN18}. These protocols have been proposed \cite{PhysRevLett.87.097901,RN47} and experimentally demonstrated \cite{RN4,RN70,RN71,RN72,RN63,RN44,PhysRevA.74.020302,PhysRevA.78.010302,sci.rep.6.19048}.
    
    These geometric phases are Abelian as the phase factors associated with distinct loops are commutable. Non-Abelian geometric phase offers an alternative strategy for realizing noise-resillient logic gates. Such phases were first discovered in a quantum system whose Hamiltonian has generate eigenstates \cite{RN21}. An adiabatically and cyclically change of such a Hamiltonian will result in a matrix-valued phase factor. Quantum logic operations based on such non-Abelian phases are referred to as adiabatic Holonomic gates \cite{RN23,RN24}. Despite fundamental interest, the practicality of such gates is also challenged by the adiabatic condition, resulting in limited experimental realization \cite{RN38,RN39}. To remove this restriction, non-adibatic holonomic gates have been proposed \cite{RN22, RN48, RN49,Chin.Sci.Bull.66.1935,phys.Rep.1027.1,Sci.chin.inf.sci.66.180502} and further improved \cite{PhysRevA.92.052302,physLettA.380.65,PhysRevA.94.052310,PhysRevA.98.052315,PhysRevA.101.062316,RN51,PhysRevLett.122.140501,PhysRevA.100.012329,RN52,PhysRevA.89.062312,PhysRevA.89.042302,PhysRevA.103.012205,RN88,RN112}. So far, such gates have been experimentally implemented in different systems, including superconducting circuits \cite{RN55,RN57,RN58,RN59,RN66,RN69,RN78,RN57,RN45,RN85}, nuclear magnetic resonance \cite{RN20,RN60,RN75}, nitrogen-vacancy centers \cite{RN56,RN61,RN62,RN64,RN65, RN68}, trapped ions \cite{RN67}.
   
    The above protocols focus on discrete-variable-encoded qubits. During the past decade, increasing efforts in quantum error correction \cite{RN86, RN87, RN88, RN96} have been paid to continuous-variable encoding schemes, where the logic qubits are encoded in some continuous-variable of bosonic modes, e.g., amplitudes of light fields \cite{RN96, RN123, RN124, RN127, RN130, RN95, RN116,RN144}. Thanks to the infinite-dimensional Hilbert space of bosonic modes, such encoding schemes are hard-ware efficient in contrast to surface code \cite{RN88,RN107,RN108,RN109,RN110,RN112}. Among various continuous-variable-encoded qubits, the cat qubit is particularly intriguing, whose codewords are encoded in two cat states with opposite parities \cite{RN294, RN94, RN89,  RN129, RN91, RN122, RN90, RN121, RN138, RN139}. The most remarkable feature of such qubits is noise bias. In other words, with the increase of cat size the bit flip error is exponentially suppressed, while the phase flip error only linearly grows \cite{RN92, RN129, lescanne2020exponential,RN294}. More importantly, such an encoding scheme may circumvent the no-go theorem, which states that it is impossible to realize a universal set of logic gates in a fault-tolerant manner with conventional encoding schemes \cite{RN92}. Up to now, dynamical logic gates for cat qubits have been proposed \cite{RN294,RN89,RN92,RN129,RN91,RN137,RN295} and demonstrated \cite{RN93, RN299}. Recently, adiabatic holonomic gates based on the encoding schemes with coherent states or cat states have been proposed \cite{RN122}, but these gates suffer from long operation time. In addition, non-adiabatic geometric gates for such qubits have been presented, but which are restricted to a small cat size \cite{RN1} or require an ancilla qutrit \cite{RN2}. We here present a scheme for realizing non-adiabatic holonomic gates with cat qubits. In addition to the non-Abelian character, the gate can overcome these problems in Refs. \cite{RN1,RN2}. In our scheme, the codewords of the qubit is encoded in two degenerate eigenstates of a  Kerr Parametric Oscillator (KPO), the even and odd cat states. By coupling these two cat states to another eigenstate with a two-photon drive and a one-photon drive, a non-adiabatic holonomic gate for the cat qubit can be realized. We further investigate the implementation of holonomic gates between two cat qubits.

    \begin{figure}
        \centering  
        \includegraphics{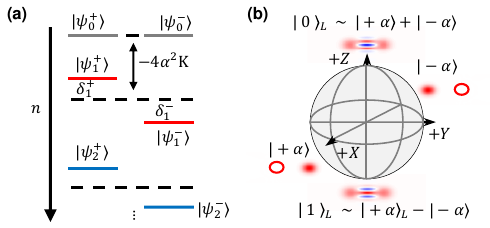}
        \caption{(a) Energy spectra of the continuous variable KPO. The eigenstates are divided into even and odd parities. $\delta_n$ is the energy gap between the $n$-th pair of eigenstates. (b) Bloch sphere of the cat qubit and the corresponding Wigner function representations (shown for $\abs{\alpha}^2 = 2.34$). The two cat states act as the logical qubits and are encoded as $\ket{0}_L$ and $\ket{1}_L$.}
        \label{fig1}
    \end{figure}
    
\section{Spectrum of the KPO system}
    We consider a KPO system, for which the cat qubit spanned by the coherent states can be confined within the subspace \{$\ket{\pm \alpha}$\} by the Kerr non-linearity and the two-photon drive. The Hamiltonian in the frame rotating at half of the oscillator frequency is
    
    \begin{equation}\label{eqKPO}
        \hat{H}_{\rm{KPO}} = -{\rm K}\hat{a}^{\dagger2}\hat{a}^2 + P(\hat{a}^2 + \hat{a}^{\dagger2}),
    \end{equation}
    where $\hat{a}$ and $\hat{a}^\dagger$ are the annihilation and creation operators for the KPO, ${\rm K}$ is Kerr nonlinearity, and $P$ is the amplitude for the drive. This Hamiltonian can be rewritten as $\hat{H}_{\rm{KPO}}=-{\rm K}(\hat{a}^{\dagger 2}-\alpha^{*2})(\hat{a}^2-\alpha^2) + P/{\rm K}$, where $\alpha=\sqrt{P/\rm K}$ , clearly showing that the two coherent states $\ket{\pm\alpha}$ are eigenstates of $\hat{H}_{\rm{KPO}}$.
    
    The energy spectra of the KPO are shown in Fig.~\ref{fig1}(a), where the eigenstates $\ket{\psi_n^\pm}$ are spanned with two different parity; wherein the even/odd cat states \added{$\ket{\psi_0^\pm} = \mathcal{N}_0^\pm(\ket{\alpha}\pm\ket{-\alpha})$} construct the degenerate eigenstate space, which can be exploited to be a couple of computational bases [Fig.~\ref{fig1}(b)]. The eigenenergies are figured out approximately by the shifted Fock basis picture transformation \cite{RN294} as
    \begin{equation}
        E_n^\pm = -4n{\rm K}\abs{\alpha^2}+\delta^\pm_n,
    \end{equation}
    where $\pm$ labels the parity of eigenstates, $n = 1,2,...$  and $\delta_n$ exponentially decreases as $\alpha$ goes up. When $\alpha$ is large, the eigenstates can be described by the shifted Fock states \added{$\ket{\psi_n^\pm} = \mathcal{N}_n^\pm[\hat{D}(\alpha) \pm (-1)^n\hat{D}(-\alpha)]\ket{n}$}.
    
\begin{figure}
        \centering
        \includegraphics{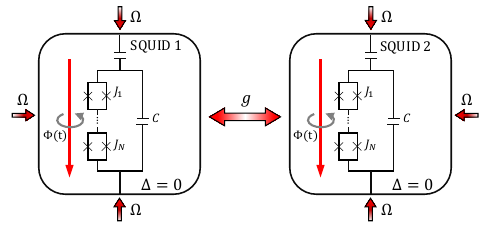}
        \caption{Circuit diagram of the coupled KPOs. The KPO in each node is controlled by a high-frequency external flux $\Phi$. The double-headed arrow denotes the coupling between two KPOs with the coupling strength $g$ and the single-headed arrows denote the external drives with strength $\Omega$.} 
        \label{fig2}
    \end{figure}

    To engineer the KPO system, we may consider the model as reported in \cite{RN142}. As shown in Fig.~\ref{fig2}, each node of the coupled KPOs consisting of a large capacitor and $N$ junctions are induced by a high-frequency magnetic flux to construct the superconducting quantum interference device (SQUID) array.
    
\section{Single-qubit gates}
    In the following, we show how to perform single-qubit non-adiabatic holonomic operations in the KPO system. As shown in Fig.~\ref{fig3}(a), the first to realize such operations is to construct a three-level configuration (see Appendix \ref{appendixA}), such as $\Lambda$ \cite{RN48}, $V$ \cite{RN61} and $\Xi$ \cite{RN69} forms. Here we choose $\{\ket{\psi_0^+},\ket{\psi_0^-}, \ket{\psi_1^+} \}$ as the bases to act as the computational subspace $\{\ket{0}_L,\ket{1}_L, \ket{2}_L \}$ for the non-adiabatic holonomic operation. Due to the potential leakage to the other states, we also consider $\ket{\psi_1^-}$ as the leaked state which is labeled $\ket{3}_L$. \deleted{Here, we select the pair of degenerate states $\ket{\psi_0^{\pm}}$ as computational bases.}
    
    \begin{figure}
        \centering  
        \includegraphics{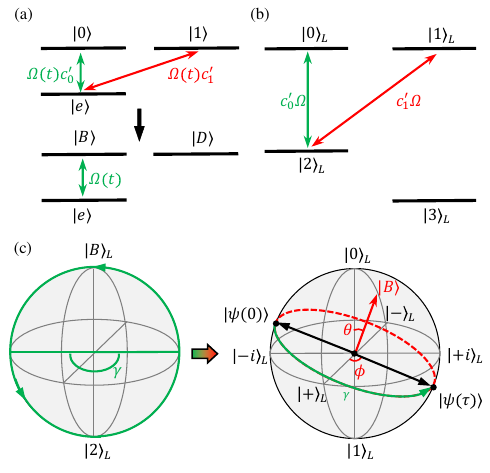}
        \caption{(a) The mechanism of non-adiabatic holonomic quantum operations in a $V$ configuration. This interaction can be equivalent to the oscillation between the bright state $\ket{B} = -a\ket{0} + b\ket{1}$ and $\ket{e}$. (b) The mechanism of the non-adiabatic holonomic evolution in \added{the} continuous variable KPO system under the single-photon and two-photon \added{drives}. The drive amplitudes here are chosen as constant for convenience. (c) Dynamics of non-adiabatic \added{holonomic} interaction in a cyclic evolution. This process can be equivalent to the rotation along axis of the bright state $\ket{B}$ with angle $\gamma$.}
        \label{fig3}
    \end{figure}
    
    In our scheme, we consider $V$ configuration to perform the non-adiabatic holonomic dynamics. We use two-photon and single-photon drives with different amplitudes $\Omega c_0/2$, $\Omega c_1/2$ but with same frequencies $\omega$. The system can be described as below:
    \begin{equation}\label{single}
    \begin{split}
        &\hat{H}_{\rm{Single}} = \hat{H}_{\rm{KPO}} + \hat{H}_{t} + \hat{H}_{s},\\
        &\hat{H}_{t} = \frac{\Omega}{2}\left[ c_0 e^{-i(\omega t+\xi)}\frac{\hat{a}^2}{2} + \rm{H.c.}\right],\\
        &\hat{H}_{s} = \frac{\Omega}{2}\left[c_1 e^{-i(\omega t+\xi)}\hat{a} + \rm{H.c.}\right],
    \end{split}
    \end{equation}
    where $\hat{H}_{t}$ and $\hat{H}_s$ represent two-photon and single-photon drives, respectively. We let $\omega = (E_1^+ - E_0)/\hbar$\added{. Under} the condition $\Omega\ll(E_j - E_0)/\hbar$ ($j=3,4,5,...$), the transitions with higher excited states can be ignored due to the rotating wave approximation \deleted{(RWA)}. Owing to parity selectivity, the single-photon drive only couples the transition $\ket{1}_L\leftrightarrow\ket{2}_L$, while the two-photon drive only couples the transition $\ket{0}_L\leftrightarrow\ket{2}_L$. In the interaction picture with respect to $H_0$, the Hamiltonian can be rewritten as:
    \begin{equation}\label{Hsingle}
        \hat{H}_{\rm{Single}}' = \frac{\Omega}{2}\left(c_0'e^{-i\xi}\ket{2}_L\bra{1} + c_1'e^{-i\xi}\ket{2}_L\bra{0} + \rm{H.c.}\right),
    \end{equation}
    where $c_0^{'} = c_0\bra{1}a\ket{2}_L$ and  $c_1^{'} =c_1 \bra{0}a^2\ket{2}_L/2$. We control $\abs{c_0'}^2+\abs{c_1'}^2=1$. The Hamiltonian of Eq. (\ref{Hsingle}) processes a dark state $\ket{D} = -c_0'\ket{0}_L + c_1'\ket{1}_L$, which remains steady during the evolution, while keeping the bright state $\ket{B} = {c_1'}^* \ket{0}_L + {c_0'}^*\ket{1}_L$ evolving through the oscillation mediated with the excited state $\ket{2}_L$. 
	
    After a cyclic evolution with the driving duration $T_g$, the whole system turns back to the computational subspace and obtains a relative phase $\gamma = \int_{0}^{T_g}\Omega/2 dt = \pi + \xi$. This process can be characterized by the transformation: $\hat{U}_{\rm{Single}} = \ket{D}\bra{D} + e^{i\gamma}\ket{B}\bra{B}$. This transformation is a universal operation, which can be regarded as a rotation \deleted{of the qubit} by an angle $\gamma$ around the axis of $\ket{B}$ on the Bloch sphere, as shown in Fig.~\ref{fig3}~(c). Provided that $\xi=0$,  $c_0'=\sin(\theta/2)e^{-i\phi}$ and $c_1'=\cos(\theta/2)$, this corresponds to $\mathbf{n}\cdot\mathbf{\sigma}$, where $\mathbf{n}=(\sin{\theta}\cos{\phi}, \sin{\theta}\sin{\phi},\cos{\theta})$ and $\mathbf{\sigma}=(\hat{\sigma}_x, \hat{\sigma}_y, \hat{\sigma}_z)$ (see Appendix \ref{appendixA}). The evolution operator for the subspace $\{\ket{0}_L, \ket{1}_L\}$ is:
    \begin{equation}
    \hat{U}_{\rm{Single}}(\theta, \phi) =
    \begin{pmatrix}
      \cos\theta& e^{-i\phi}\sin\theta\\
        e^{i\phi}\sin\theta & -\cos\theta 
    \end{pmatrix}.
    \end{equation} 
    By controlling $\xi$, $c_0$ and $c_1$, we can realize the universal single-qubit gates with different parameter $(\gamma, \phi, \theta)$. Since the parallel-transport condition $ _L\bra{k}\hat{H}_{\rm{Single}}^{'}\ket{l}_L=0$ ($k,l=0,1$) is satisfied, this evolution is purely geometric. What's more, it is also a non-Abelian operation and can be verified with the choices of $(\gamma,\theta,\phi) = \{(0, 0, 0), (0, 0,\pi/2) \}$, which correspond to $\hat{\sigma}_z$ and $\hat{\sigma}_x$ operations, respectively.  We perform these two operations on arbitrary quantum states in a different order and test to verify that the results of final states are promising.
    
   \begin{figure}
        \centering  
        \includegraphics[width=8.3cm]{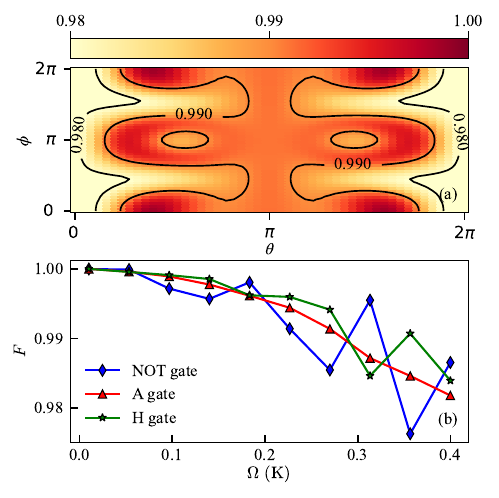}
        \caption{\added{Simulation results of the single-qubit gates. (a) The gate fidelity, defined in Eq. (\ref{eqfqpt}), versus the parameters ($\theta$, $\phi$) within $[0, 2\pi]$ for $\Omega = 0.25{\rm K}$ and $\gamma=0$. (b) The gate fidelity of the NOT, A and H gates as a function of $\Omega/{\rm K}$, with ($\abs{\alpha}^2$,  $T_g$) = (2.3,  $2\pi/\Omega$). The parameters ($\theta$, $\phi$) for the simulated NOT, A and H gates are chosen as ($3\pi/2$, 0), ($3\pi/2$, $\pi/4$) and ($7\pi/4$, 0), respectively.}}
        \label{fig4}
    \end{figure}
    
    We now turn to the discussion of \added{the gate operation time}, which is limited by the Kerr nonlinearity ${\rm K}$. In general, a large ${\rm K}$ is favorable. \deleted{(}As shown below, the gate operation time is proportional to ${\rm K}^{-1}$ \cite{RN295}.\deleted{)} \added{To verify the above discussion, we numerically calculate the gate fidelity through the method of the quantum process tomography, which is defined as \cite{RN69, ning2019deterministic} 
    \begin{equation}\label{eqfqpt}
        F = \tr(\chi_{\rm id}\chi_{\rm num}),
    \end{equation}
    where $\chi_{\rm id}$ is the ideal process matrix and $\chi_{\rm num}$ is the process matrix obtained by the numerical simulation. The parameters are set as $\Omega=0.25{\rm K}$ and $\abs{\alpha}^2=2.34$. }
    
    \added{As shown in Fig.~\ref{fig4}(a), the gate operations with partial angles have relatively low fidelities, which is due to incomplete gate operations and the leakage to the higher excited states.} This error can be reduced to the minimum through the control of $c_0$ and $c_1$. Actually,  arbitrary non-adiabatic holonomic single-qubit gates \added{for} parameters ($\theta$, $\phi$) can be achieved with high fidelity as long as the Rabi frequency of the drive $\Omega$ is controlled appropriately. The gate fidelity goes up as $\Omega$ decreases, but the corresponding time  $T_g = 2\pi/\Omega$ will be extended. Notice that $\Omega$ is still required to be optimized so as to completely avoid the coupling of the computational space to the higher excited states.
    
    \begin{figure}
        \centering  
        \includegraphics[width=8.3cm]{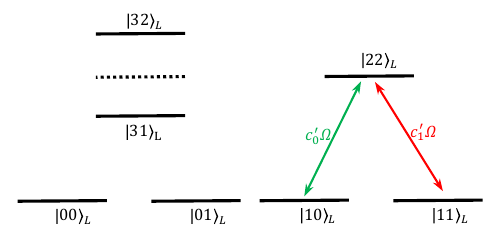}
        \caption{Partial level construction of the two-qubit KPO system, $\{\ket{00},\ket{01},\ket{10},\ket{11}\}$ are degenerate ground states and $\ket{22}$ is the excited state. Here we consider $\{\ket{31},\ket{32}\}$ as leakage states due to level leakage during gate operation process.}
        \label{fig5}
    \end{figure}
    If we choose small $\alpha$, all the energy \added{gaps} are too small to cause leakage to higher excited states but not only the excited state $\ket{3}_L$. When $\alpha$ is too large, the leakage to $\ket{3}_L$ will be intensified, scrambling evolution of the non-adiabatic holonomic process. Maybe this limit can be overcome through derivative-based corrections \deleted{(DBC)} \cite{RN294}. 
    
    To avoid the effect of leakage, we assume $\alpha^2$ is large enough so that we can choose an appropriate $\Omega$ to perform non-adiabatic holonomic evolution. Similar to those reported in Ref.~\cite{RN61}, we choose $(\theta,\phi)$ as $(3\pi/2, 0)$, $(3\pi/2, \pi/4)$, and (7$\pi$/4, 0), which correspond to NOT, A, and H gates, respectively. These gates constitute a complete set of single-qubit \added{operations}. The gate fidelities with $\Omega$ are shown in Fig.~\ref{fig4}(b). It can be found that, the fidelities of these gate become higher as the amplitude of drive increases, and the operation time decreases as the Kerr nonlinearity increases. 
 
\section{Two-qubit gate}
    To realize a two-qubit non-adiabatic holonomic gate, we also need to construct \added{a} $V$ model, whose realization depends on the control-qubit: only the control-qubit is in $\ket{1}_L$ \added{such} that the target-qubit flips through the holonomic dynamics otherwise remains steady. The Hamiltonian can be written as follows:
    \begin{equation}
        \hat{H} = \ket{1}_L\bra{1}\otimes \hat{H}'_{\rm{Single}} + \ket{0}_L\bra{0}\otimes \hat{I},
    \end{equation}
    where $\hat{I}$ is identity operator of the cat qubit, and $\hat{H}'_{\rm{Single}}$ is the single-qubit Hamiltonian in Eq. (\ref{Hsingle}). Here we induce two non-energy-conserving couplings to realize the two-qubit gate.
    \begin{figure}
        \centering  
        \includegraphics[width=8.3cm]{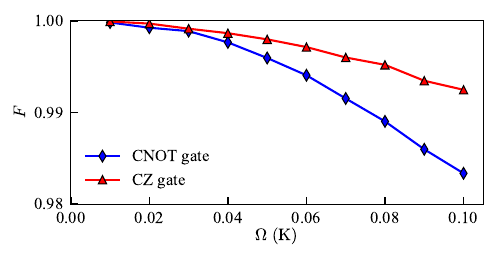}
        \caption{\added{Simulated gate fidelities, defined in Eq.~(\ref{eqfqpt}), of the CNOT and CZ gates versus $\Omega/{\rm K}$, with ($\abs{\alpha}^2$,  $T_g$) = (2.3,  $2\pi/\Omega$). The parameters ($\theta$, $\phi$) for the simulated CNOT and CZ gates are chosen as ($3\pi/2$, 0) and (0, 0), respectively.}}
        \label{fig6}
    \end{figure} 
    
    As shown in Fig.~\ref{fig5}, the two drives couple $\ket{10}_L$ and $\ket{11}_L$ to $\ket{22}_L$ but not cause the coupling of $\ket{00}_L$ and $\ket{01}_L$ to any other states ($\ket{jk}_L \equiv \ket{j}_L \otimes \ket{k}_L$, the 1st and 2nd vectors represent the control qubit and the target qubit), due to parity selectivity and large detuning. The Hamiltonian is:
    \begin{equation}
        \hat{H}_{\rm{Two}} = \hat{H}_{\rm{KPO},2} + \hat{H}_{t,2} + \hat{H}_{s,2},
    \end{equation}
    with
    \begin{equation}
        \begin{split}
            &\hat{H}_{\rm{KPO},2} = \hat{H}_{\rm{KPO}}\otimes \hat{I} + \hat{I} \otimes \hat{H}_{\rm{KPO}},\\
            &\hat{H}_{t,2} = \frac{\Omega}{2}\left[c_0 e^{-i(\omega_{22} t+\xi)}\hat{a}\otimes \hat{b}^2+\rm{H.c.}\right],\\
            &\hat{H}_{s,2} = \frac{\Omega}{2}\left[c_1 e^{-i(\omega_{22} t+\xi)}\hat{a}\otimes \hat{b} + \rm{H.c.}\right],
        \end{split}
    \end{equation}
    where $\hat{H}_{\rm{KPO},2}$ describes the two-qubit KPO system, $\hat{a}$ and $\hat{b}$ are the annihilation operators for the control-KPO and the controlled-KPO, respectively, $\hat{H}_{t,2}$ represents two-photon coupling for \added{the} $\ket{10}\leftrightarrow\ket{22}$ transition and $\hat{H}_{s,2}$ is single-photon coupling \added{for} the $\ket{11}\leftrightarrow\ket{22}$ transition. Let $\xi=0$\added{. After} a cyclic evolution, the corresponding evolution operator of the cat qubits is:
    \begin{figure*}
    \centering  
    \includegraphics{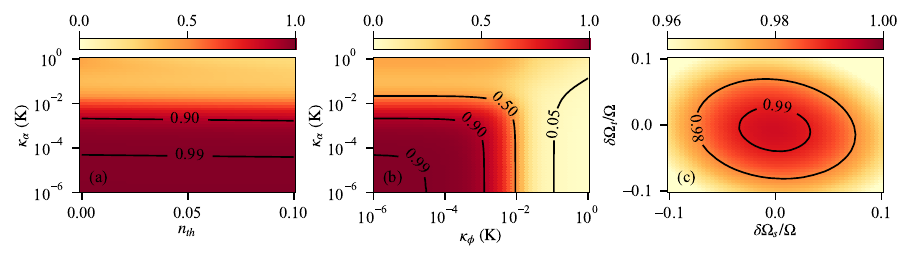}
    \caption{\added{The simulated results for the fidelity, defined in Eq.~(\ref{eqfqpt}), of the NOT gate under the influences of noises, for $\Omega=0.23{\rm K}$. (a) The gate fidelity as a function of rate of the rescaled single-photon loss $\kappa_\alpha/{\rm K}$ and the number of thermal photons $n_{t h}$. (b) The gate fidelity as a function of the rescaled single-photon loss rate $\kappa_\alpha/{\rm K}$ and the rescaled dephasing rate $\kappa_{\phi}/{\rm K}$. (c) The gate fidelity as a function of the rescaled amplitude fluctuation for the single-photon drive, $\delta\Omega_s/\Omega$, and for the two-photon drive, $\delta\Omega_t/\Omega$.}}
    \label{fig7}
    \end{figure*} 
    \begin{equation}
        \hat{U}_{\rm{Two}}(\theta, \phi) =\begin{pmatrix}
        1&0&0&0\\
        0&1&0&0\\
        0&0&\cos\theta & e^{-i\phi}\sin\theta\\
        0&0&e^{i\phi}\sin\theta&-\cos\theta
    \end{pmatrix}.
    \end{equation}
    We can realize the CNOT gate with the choice of \added{$(\phi,\theta) = (0,3\pi/2)$} and the CZ gate with the choice of \added{$(\phi,\theta) = (0,0)$}. To verify this method, we perform the numerical simulation. As can be seen from Fig.~\ref{fig6}, the CNOT gate with fidelity above $99\%$ can be achieved for \added{$\Omega<0.08{\rm K}$}, while the case for the CZ gate is better, with the main error due to the leakage out of the \added{non-adiabatic} holonomic state evolution subspace. In fact, the constructed gate operation can be extended to non-adiabatic non-Abelian arbitrary two-qubit gate operation.
\section{Analysis of decoherence and leakage}
    In this section, we study the impact of decoherence process and leakage to the higher excited states. The full \added{dynamics of the driven KPO} can be described by the Markov master equation \cite{noise}:
    \begin{equation}
    \begin{split}
        \frac{d\hat{\rho}}{d t}=&-i[\hat{H}_{\rm{Single}},\hat{\rho}]+\kappa_\alpha(1+n_{t h})D[\hat{a}]\hat{\rho}\\
        &+\kappa_\alpha n_{t h}D[\hat{a}^\dagger]\hat{\rho} +\kappa_\phi D[\hat{a}^\dagger\hat{a}]\hat{\rho},
    \end{split}
    \end{equation} 
    where $\kappa_\alpha$ is the single-photon loss rate, $n_{th}$ is the number of thermal photons and $\kappa_\phi$ is the rate of pure dephasing. We analyze it from two aspects: (1) the gain/loss of the photon and (2) the dephasing, with the method in \cite{RN93}, respectively. 
    
    As mentioned above, the first pair of excited states can be written by the shifted Fock states \added{$\ket{\psi_n^{\pm}}=\mathcal{N}_n^\pm[\hat{D}(\alpha)+(-1)^{\pm}\hat{D}(-\alpha)]\ket{1}$}. In this approximation, the \added{actions} of  $\hat{a}$ and $\hat{a}^{\dagger}$ have the following effects:
	\begin{equation}
	\begin{split}
		&\hat{a}\ket{\psi_0^{\pm}}=\alpha \ket{\psi_0^{\mp}},\\
		&\hat{a}^\dagger\ket{\psi_0^{\pm}}=\alpha \ket{\psi_0^{\mp}}+\ket{\psi_1^{\mp}}.
	\end{split}
	\end{equation}
	
	These two operators, corresponding to single-photon loss and thermal photon production process, will cause bit flip and leakage of cat qubit bases, respectively. In Fig.~\ref{fig7}(a), we show the fidelity of the NOT gate as functions of $\kappa_\alpha$ and $n_{t h}$ with $\Omega=0.23{\rm K}$, corresponding to $T_g\approx27{\rm K}^{-1}$. This result is similar to those of H and A gates, respectively. It can be found that the \added{influences} of the thermal photons increases as the rate of single-photon loss \added{increase}. In fact, when $\kappa_\alpha/8{\rm K}\abs{\alpha}^2\ll1$, the \added{influences} of the single-photon loss \added{cause} phase space distortion of cat-qubits $\ket{\pm\alpha}\rightarrow\ket{\pm\tilde{\alpha}}=\ket{\pm r_0e^{i\theta_0}}$ \cite{RN91}:
	\begin{equation}
	\begin{split}
	    r_0&=(\frac{4P^2-\kappa_\alpha^2/4}{4{\rm K}^2})^{1/4},\\ \tan{2\theta_0}&=\frac{\kappa_\alpha}{\sqrt{16P^2-\kappa_\alpha^2}}.
	\end{split}
	\end{equation}
    The \added{influences} of dephasing can be divided into two parts: $\hat{Z}-$error and leakage to higher excited states. The first part can be suppressed by large $\abs{\alpha}^2$, but the second part is inevitable. As shown in Fig.~\ref{fig7}(b), the \added{influences} of dephasing \added{are} larger than \added{those} of single-photon loss due to the leakage to higher excited states. To perform the \added{non-adiabatic} holonomic gate, the key is to decrease leakage to higher excited states, which may be solved by combining other technologies, such as derivative-based transition suppression technique \cite{RN296}, to alleviate this noise\added{,} or to improve the Kerr nonlinearity ${\rm K}$.
    
    Finally, we study the influences of the fluctuation of the amplitudes for both the single- and two-photon drives. The gate's sensitivity to the fluctuations of  $\Omega_s=\Omega c^{'}_1$ and $\Omega_t=\Omega c^{'}_0$ is shown in Fig.~\ref{fig7}(c) with $\Omega=0.23{\rm K}$. This result shows that gate fidelity is only reduced by about $1\%$ for \added{$4\%$} of the deviation of $\Omega_s$ and $\Omega_t$, revealing the potential advantage of \added{the gate}. 
    
    Extension of numerical calculations including decoherence and leakage to the cases of the two-qubit gates is intuitive and similar, results of which are not shown due to the large overhead of our computation capacity at hand.
    
\section{\added{Discussion} and Outlook}
    In conclusion, we have shown that the non-adiabatic holonomic single- and two-qubit operations can be achieved with cat qubits through the KPO and the coupled KPOs with the combination of the single- and two-photon driving processes. Further improved gate fidelities may be reached with the support of the developed optimized methods, such as derivative-based transition suppression \added{techniques \cite{RN294}}. Experimental demonstration of these constructed gates can be expected\added{,} attributed to the \added{developments} of the single \added{cat} qubit gates \added{with a} KPO\added{,} realized with superconducting circuit systems \cite{RN93}.
\begin{acknowledgments}
    This work was supported by the National Natural Science Foundation of China under Grand Nos. 12274080 and 11875108.
\end{acknowledgments}




\appendix
\section{\label{appendixA}Non-adiabatic holonomic operation based on cat-qubit}
In this section, we derive the realization of non-adiabatic holonomic \added{operations} in the KPO system. we consider the method in \added{Ref.~\cite{RN48}} and derive this formula from Schr$\ddot{\rm{o}}$dinger equation. In the rotating frame of $\hat{H}_{\rm{KPO}}$, due to the parity selectivity of \added{the cat qubit}, the Hamiltonian in Eq. (\ref{single}) can be rewritten as:
\begin{widetext}
    \begin{equation}
        \begin{split}
        \hat{H}_r = &e^{i\hat{H}_{\rm{KPO}}t/\hbar}\hat{H}_{\rm{Single}}e^{-i\hat{H}_{\rm{KPO}}t/\hbar}\\
        =&\frac{\Omega}{2}\left[c_0 e^{-i\xi} \sum_{i,j=0}(\bra{\psi_{i}^{+}}\frac{\hat{a}^2}{2}\ket{\psi_{j}^{+}}e^{i\delta_{ij}^{++}t}\ket{\psi_{i}^{+}}\bra{\psi_{j}^{+}}+\bra{\psi_{i}^{-}}\frac{\hat{a}^2}{2}\ket{\psi_{j}^{-}}e^{i\delta_{ij}^{--}t}\ket{\psi_{i}^{-}}\bra{\psi_{j}^{-}})+ \rm{H.c.}\right] \\
        &+\frac{\Omega}{2}\left[c_1e^{-i\xi}\sum_{i,j=0}(\bra{\psi_{i}^{+}}\hat{a}\ket{\psi_{j}^{-}}e^{i\delta_{ij}^{+-}t}\ket{\psi_{i}^{-}}\bra{\psi_{j}^{+}}+\bra{\psi_{i}^{-}}\hat{a}\ket{\psi_{j}^{+}}e^{i\delta_{ij}^{-+}t}\ket{\psi_{i}^{+}}\bra{\psi_{j}^{-}}) + \rm{H.c.}\right],
        \end{split}
    \end{equation}
\end{widetext}
where $\delta_{ij}^{+-}=(E_i^+ - E_j^-)/\hbar-\omega$ and $E_0^+=E_0^-$. We control $\omega = [E_1^+-E_0^+({\rm or} \ E_0^-)]/\hbar$ and $\Omega\ll \abs{[E_1^--E_0^+({\rm or} \ E_0^-)]/\hbar}$, and drop the high-frequency oscillation terms according to \added{rotating wave approximation}. The Hamiltonian can be simplified as below:
\begin{equation}\label{Hssingle}
    \hat{H}_{\rm{Single}}^{'} = \frac{\Omega}{2}\left(c_0^{'}e^{-i\xi}\ket{2}_L\bra{1} + c_1^{'}e^{-i\xi}\ket{2}_L\bra{0} + \rm{H.c.}\right),
\end{equation}
where $c_0^{'} = c_0\bra{1}a\ket{2}_L$ and  $c_1^{'} =c_1 \bra{0}a^2\ket{2}_L/2$\added{. We} encode $\ket{\psi_0^+}$, $\ket{\psi_0^-}$\added{, }and $\ket{\psi_1^+}$ as $\ket{0}_L$, $\ket{1}_L$\added{, and} $\ket{2}_L$, respectively. Provided $\xi=0$, $c_0^{'}=\sin(\theta/2)e^{-i\phi}$ and $c_1{'}=\cos(\theta/2)$, \added{the corresponding dark state of the Hamiltonian in Eq.~(\ref{Hssingle}) is written as} $\ket{D} = -\sin(\theta/2)e^{-i\phi}\ket{0}_L + \cos(\theta/2)\ket{1}_L$\added{, }that remains steady. The \added{unitary operator of the evolution dominated by the Hamiltonian (\ref{Hssingle})} in the subspace $\{\ket{D}, \ket{B}, \ket{2}_L\}$ \added{can be modelled as}:
\begin{equation}
\hat{U}_{\rm{Single}}(T_g,0) = 
\begin{pmatrix}
1 &0&0\\
0&\cos(\Omega T_g/2)&-i\sin(\Omega T_g/2)\\
0&-i\sin(\Omega T_g/2)&\cos(\Omega  T_g/2)
\end{pmatrix},
\end{equation}
\added{which reduces to $\hat{U}_{\rm{Single}}(T_g,0) = -(\ket{2}_L\bra{2} + \ket{B}_L\bra{B}) + \ket{D}_L\bra{D}$,} with $T_g=2\pi/\Omega$. If the KPO is confined within the subspace $\{\ket{0}_L, \ket{1}_L\}$, the unitary transformation can be described as below:
\begin{equation}
    \hat{U}_{\rm{Single}}(\theta, \phi) =
    \begin{pmatrix}
    \cos\theta& e^{-i\phi}\sin\theta\\
    e^{i\phi}\sin\theta & -\cos\theta
\end{pmatrix}.
\end{equation} 
With the \added{choices} of ($\theta$, $\phi$) as ($3\pi/2$, $0$), ($3\pi/2$, $\pi/4$) and ($7\pi/4$, $0$), we can realize NOT, A, \added{and} H gates, respectively, which form a complete set of single-qubit gates.

Two-qubit \added{operations} can be done in a similar way\deleted{, but has more complicated energy spectrum and harsh constraint to avoid high level leakage}. We design the specific coupling mechanism to drive the oscillation between $\ket{B}$ and $\ket{2}_L$ of the controlled-KPO when the state of the control-KPO is $\ket{1}_L$ by the parity selectivity. To alleviate the leakage to the other excited states, we should control the coupling strength much smaller than the gap between the other excited states and the state to be driven. The process can be described as:
\begin{equation}
    \hat{U}_{\rm{Two}}(\theta, \phi) =\begin{pmatrix}
    1&0&0&0\\
    0&1&0&0\\
    0&0&\cos\theta & e^{-i\phi}\sin\theta\\
    0&0&e^{i\phi}\sin\theta&-\cos\theta
\end{pmatrix}.
\end{equation}
\bibliography{reference}

\end{document}